\documentstyle[12pt,epsf]{article}

\newcommand{\eq}{\begin{equation}}
\newcommand{\eqn}{\begin{displaymath}}
\newcommand{\en}{\end{equation}}
\newcommand{\enn}{\end{displaymath}}

\def\pnot{\mbox{${\not{\hbox{\kern-3.0pt$p$}}}$}}
\def\qnot{\mbox{${\not{\hbox{\kern-2.0pt$q$}}}$}}
\def\enot{\mbox{${\not{\hbox{\kern-2.0pt$e$}}}$}}
\def\knot{\mbox{${\not{\hbox{\kern-2.0pt$k$}}}$}}

\def\fun#1#2{\lower3.6pt\vbox{\baselineskip0pt\lineskip.9pt\ialign
{$\mathsurround=0pt#1\hfil##\hfil$\crcr#2\crcr\sim\crcr}}}

\begin{document}
\begin{titlepage}

\hskip 12cm \vbox{\hbox{DFPD 97/TH/}\hbox{CS-TH 3/97}
\hbox{December 1997}}
\vskip 0.3cm
\centerline{\bf POMERON NON FACTORIZATION}
\centerline{\bf IN DIFFRACTIVE SCATTERING AT HERA$^{~\ast}$}
\vskip 1.0cm
\centerline{  R. Fiore$^{a\dagger}$, A. Flachi$^{a,d\sharp}$, 
L. L. Jenkovszky$^{b\ddagger}$, F. Paccanoni$^{c\diamond}$}
\vskip .5cm
\centerline{$^{a}$ \sl  Dipartimento di Fisica, Universit\`a della Calabria,}
\centerline{\sl Istituto Nazionale di Fisica Nucleare, Gruppo collegato di
Cosenza}
\centerline{\sl Arcavacata di Rende, I-87030 Cosenza, Italy}
\vskip .5cm
\centerline{$^{b}$ \sl  Bogoliubov Institute for Theoretical Physics,}
\centerline{\sl Academy of Sciences of the Ukrain}
\centerline{\sl 252143 Kiev, Ukrain}
\vskip .5cm
\centerline{$^{c}$ \sl  Dipartimento di Fisica, Universit\`a di Padova,}
\centerline{\sl Istituto Nazionale di Fisica Nucleare, Sezione di Padova}
\centerline{\sl via F. Marzolo 8, I-35131 Padova, Italy}
\vskip .5cm
\centerline{$^{d}$ \sl Department of Physics, University of Newcastle upon 
Tyne}
\centerline{\sl Newcastle upon Tyne NE1 7RU, United Kingdom}
\vskip 1cm
\begin{abstract}
Pomeron non factorization for the diffractive scattering is considered and 
an analysis of experimental data is performed with different cuts in the 
momentum fraction $\xi$. The results of the analysis are compared with the 
breaking predicted in the Genovese-Nikolaev-Zakharov model. 
\end{abstract}
\vskip .5cm
\hrule
\vskip.3cm

\noindent
$^{\diamond}${\it Work supported in part by the Ministero italiano
dell'Universit\`a e della Ricerca Scientifica e Tecnologica and in part
by INTAS}
\vfill
$\begin{array}{ll}
^{\dagger}\mbox{{\it email address:}} &
   \mbox{FIORE ~@CS.INFN.IT}
\end{array}
$

$\begin{array}{ll}
^{\sharp}\mbox{{\it email address:}} &
   \mbox{antonino.flachi~@ncl.ac.uk}
\end{array}
$

$ \begin{array}{ll}
^{\ddagger}\mbox{{\it email address:}} &
 \mbox{JENK~@GLUK.APC.ORG}
\end{array}
$

$ \begin{array}{ll}
^{\diamond}\mbox{{\it email address:}} &
   \mbox{PACCANONI~@PADOVA.INFN.IT}
\end{array}
$
\end{titlepage}
\eject
\textheight 210mm
\topmargin 2mm
\baselineskip=24pt

{\bf 1. Introduction}

Factorization theorems ~\cite{CSS} have been proven, order by
order in perturbative QCD, for totally inclusive cross-sections.
Factorization of short- and long-distance effects in the hadron-
hadron scattering provides all possible corrections to the 
parton model predictions at high energy. For the diffractive scattering,
when one of the initial hadrons changes only its momentum,
the factorization theorem fails. The amount of the 
factorization breaking depends strongly
on the kinematical region of the considered process. The case
of coherent hard diffraction, where a heavy quark or a jet is
produced in addition to the diffraction, has been considered
in detail ~\cite{FS,CFS}. In this case, non-factorization
occurs at the lowest relevant order of perturbation theory
~\cite{CFS}.

If we limit ourselves to diffractive DIS at HERA, we can neglect
the factorization breaking due to the coherent Pomeron
~\cite{CFS} and explore different aspects of the problem.
According to a conventional wisdom, the diffractive scattering at 
high energy is dominated by the Pomeron exchange.
The Ingelman-Schlein model ~\cite{IS} imposes factorization 
also for semi-inclusive diffractive processes and applies to the 
hadron-Pomeron scattering. The Pomeron is 
considered as a bound state of gluons and quarks that interact
with hadrons, or leptons in deep inelastic scattering (DIS), and
when the modulus of the squared momentum transfer, $-t$, is
small the picture sketched above seems appropriate.

The events we consider in the following are characterized by
a large rapidity gap between the unscathed proton and the
other final state particles ~\cite{TA,MD}. From now on, factorization
and its breaking refer more to the factorization property of
the residues of the Regge pole exchanged than to the
separation of the hard from the soft scattering in the 
process. This form of factorization has been referred to as 
"gap factorization" ~\cite{KG}. Since the
exchange of a meson gives also rise to rapidity gaps, a 
natural explanation for the breakdown of gap factorization,
observed experimentally ~\cite{H1,H2}, can be found in the
exchange of different trajectories ~\cite{DL}. Different structures of 
the Pomeron and mesons and interference terms (that must be
present in the diffractive structure function since it is
a cross section) break the factorization property ~\cite{All,KG}. 
Factorization breaking, as it appears experimentally ~\cite{H1}
and as interpreted with the above mechanism, is limited to
values of $\xi$, the momentum fraction lost by the proton,
larger than $\approx 0.005$. In the dipole approach ~\cite{GNZ,NZ}
to the BFKL Pomeron ~\cite{BFKL}, instead, the failure of
factorization is a phenomenon involving also a pure Pomeron
exchange and is present for all values of $\xi$. The microscopic
dipole BFKL Pomeron has a two-component structure function and
each component has a different coupling to the proton.
To each piece, in fact, a different flux of Pomerons in the
proton is associated that distinguish between the "valence
$q \bar{q}$" and the "valence gluons plus sea" component.
A similar effect has been found in Ref.~\cite{WU}.

From the published data ~\cite{TA,MD} it is difficult to draw
definite conclusions about the Pomeron non-factorization. In
order to suppress the contribution from meson trajectories,
only data with $\xi \leq 0.005$ must be selected and this
reduces the already limited statistics of the available data.
However it could be possible to distinguish different trends
with and without a cut in the $\xi$ variable. From a
theoretical point of view it is possible to show that,
under quite reasonable conditions, a small 
departure from factorization of the
Pomeron exchange is present  already in the Ingelman-Schlein
model if the $t$-dependence of the
diffractive structure function is taken into account.

In this paper we consider the above points in a Regge 
model of the diffractive scattering. After a brief discussion of
the kinematics, we present an analysis of the experimental data,
where different cuts in $\xi$ lead to quite different behaviours
of the Pomeron intercept. We give the proof that, by integrating over
$t$ the 4-variable diffractive structure function, the Pomeron
factorization can be lost and compare this effect with the
breaking predicted in Ref.~\cite{GNZ}. Finally we draw the conclusions.

\vskip 0.5cm

{\bf 2. Notations and the definition of the $\xi$-slope from experiments}

The notation and kinematics for the process
\begin{displaymath} 
e^-(l)+p(p) \to e^-(l')+p(p')+X_n(k_n)
\end{displaymath}
are shown in Fig. 1, where $r$ is the four-momentum of the
exchanged Reggeon. Then the spin averaged differential
cross section for a diffractive DIS, ignoring spin and the
proton mass, is
\begin{displaymath} 
d\sigma=\frac{(2\pi)^{-5}}{2p\cdot l}\frac{d^3\vec{l}'}{2l'^0}
\frac{d^3\vec{p}\,'}{2p'^0}
\end{displaymath}
\begin{equation}
\times \sum_n \left(\frac{d^3\vec{k}_n}{2k_n^0} \delta(p+l-l'-k_n-p')
|T(p+l\to p'+l'+k_n)|^2 \right)~,
\label{z1}
\end{equation}
where $T$ is the transition amplitude.

\begin{figure}
\begin{center}
\leavevmode
\epsfxsize=0.45\textwidth
\epsffile{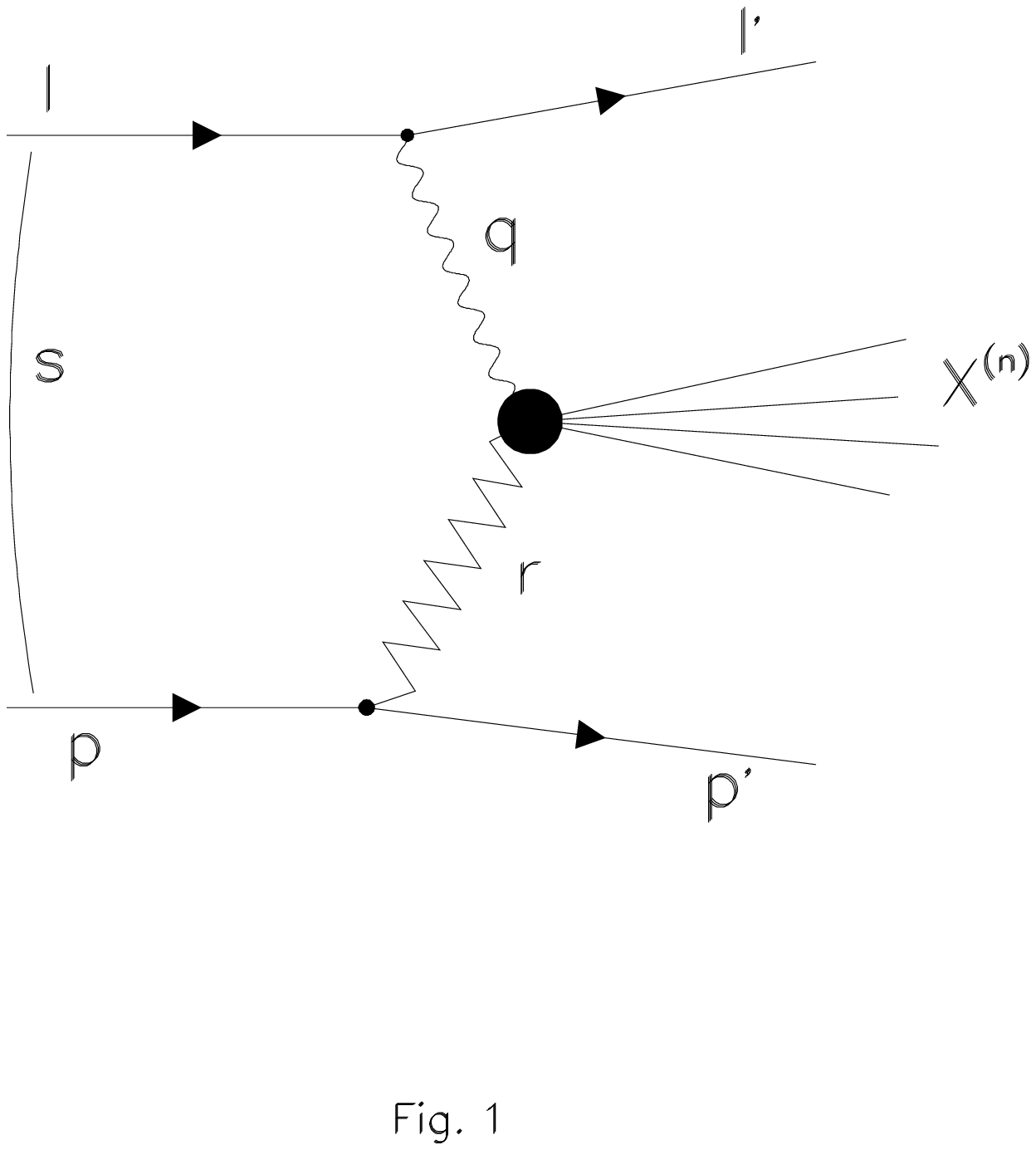}
\end{center}
\caption{\small The diagram for the diffractive deep-inelastic scattering.}
\label{fig1}
\end{figure}
          
If a sufficiently small cut in the variable $\xi$, with $\xi =
(r\cdot q)/(p\cdot q)$, is selected, then only the Pomeron
trajectory will contribute to the process and $r=p-p'$
represents, in this case, the four-momentum of the Pomeron. 
In the following, $r^2 \equiv t <0$ is the squared mass of
the Pomeron.

Gap factorization implies that
\begin{equation}
T(p+l\to p'+l'+k_n)=F(l+r\to l'+k_n) \Phi(\xi,t)~,
\label{z2}
\end{equation}
where the "flux factor" $\Phi(\xi,t)$ has the form
\begin{equation}
\Phi(\xi,t) =(e^{i\pi /2}\xi)^{-\alpha_P(t)}\beta(t)~,
\label{z3}
\end{equation}
$\alpha_P(t)$ being the Pomeron trajectory. Since only the
Pomeron is exchanged, Eq.~(\ref{z1}) becomes
\begin{displaymath}
d\sigma =\frac{r\cdot l}{p\cdot l}\frac{d^3\vec{p}\,'}{2p'^0}
|\Phi(\xi,t)|^2 d\sigma(l+r\to l'+k_n) \simeq
\end{displaymath}
\begin{equation}
\frac{\pi}{2}\xi\,d\xi\,dt |\Phi(\xi,t)|^2 d\sigma(l+r\to
l'+k_n)~,
\label{z4}
\end{equation}
where the relations $(r\cdot l)/(p\cdot l) \sim (r\cdot q)/
(p\cdot q) = \xi$ and $d^3\vec{p}\,'/(2p'^0)\propto d\xi\,dt$,
valid for small $\xi$, were used.

Let now $k$ be the four-momentum of the parton interacting with 
the lepton and $\beta$ the momentum fraction of the Pomeron
carried by the parton:
\begin{displaymath}
k=\beta r~.
\end{displaymath}
Then, with the above mentioned approximations, one may write
$(Q^2=-q^2)$
\begin{equation}
d\sigma(l+r)=G_{q/P}(\beta,Q^2,t)\frac{d\hat{\sigma}(k+l)}{dq^2}
d\beta\,dq^2
\label{z5}
\end{equation}
in terms of the parton-lepton cross section $d\hat{\sigma}$,
where $G_{q/P}(\beta,Q^2,t)$ is the structure function of the
Pomeron. By introducing the 4-variable diffractive structure function
$F_2^{D(4)}(\beta,Q^2,\xi,t)$ we can rewrite Eqs.~(\ref{z4}) and (\ref{z5}) 
as
\begin{equation}
F_2^{D(4)}= \xi\,|\Phi(\xi,t)|^2 G_{q/P}(\beta,Q^2,t)~.
\label{z6}
\end{equation}
We may consider two rapidity gaps in the whole process: the large 
rapidity gap determined by $\ln(1/\xi)$ and the gap $\ln(1/\beta)$
involved in the creation of that quark from the Pomeron, 
which will finally interact with the photon. For small $\beta$ the
squared mass $k_n^2$ of the produced hadrons, besides the proton,
is large and the triple Pomeron vertex will play an essential
role. For large $\beta$, instead, the above formalism applies if
the quark of four-momentum $k$ is near its mass-shell. The
corresponding $t$-dependence of the Pomeron structure function
could be quite different in the two $\beta$ regions. This pont
is made clear in the calculation of Ref.~\cite{DL} where the three
pieces contributing to the Pomeron structure function have 
different $t$ and $\beta$ dependences. Physically they correspond
to the quark loop, i.e. the quark diagram important at large $\beta$,
the triple Pomeron vertex and the $PPf$ term.

That the Pomeron structure function must depend on the variable 
$t$ has been emphasized already in earlier papers ~\cite{GB,CAP}, where
this $t$-dependence ultimately disappears since the process is
considered near $t=0$. As we will show later, the $t$-dependence
can cause factorization breaking when Eq.~(\ref{z6}) is 
integrated over $t$ to get $F_2^{D(3)}(\beta,Q^2,\xi)$.

We consider now the experimental data.
H1 data ~\cite{TA,H1} cover a $\xi$-interval where both the Pomeron
and meson trajectories contribute. The analysis of Ref.~\cite{H1} shows
that the effective power $n$ of $\xi$ in $F_2^{D(3)}$ depends on
$\beta$. This factorization breaking (Eq.~(\ref{z6}) does not
hold anymore at $t=0$) is consistent with a sizable contribution
to $F_2^{D(3)}$ from meson exchange ~\cite{H1,All}. In 
Fig.2 this result has been reproduced, with open circles,
together with a quadratic fit for $n(\beta)$ (continuous line).

\begin{figure}
\begin{center}
\leavevmode
\epsfxsize=0.45\textwidth
\epsffile{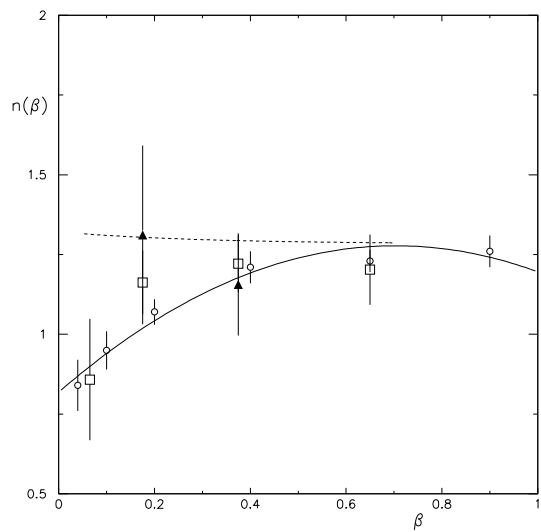}
\end{center}
\caption{\small The parameter n versus $\beta$. Open circles represent the 
result of Ref.~[8] while the continuous line comes from a quadratic fit. 
Open squares and full triangles represent the result according 
Eq.~(\ref{z7}), without and with the cut in $\xi$ respectively.}
\label{fig2}
\end{figure}

In the above analyses ~\cite{H1,All} the meson contribution dies
out for $\xi \leq 0.005$ leaving only the Pomeron, as given
in Eq.~(\ref{z6}) in a Regge model. Should the effective slope
still depend on $\beta$ in the small-$\xi$ selected sample, then
a different interpretation, as for example the one
proposed in Ref.~\cite{GNZ},
could be appropriate. A first hint to the effect we propose to
exploit can be found in the published H1 ~\cite{TA} and
ZEUS ~\cite{MD} data. Due to the different explored $\xi$-range,
we can say that the ZEUS data are essentially determined from 
Pomeron exchange while H1 data are not. The mean value of the
$\xi$ exponent, the same for each $\beta$ and $Q^2$ bin, is
larger for ZEUS $(\sim 1.3)$ than for H1 $(\sim 1.19)$.
In a recent analysis ~\cite{ZEUS}, with a larger kinematic range
$4\times 10^{-4}<\xi <0.03$, ZEUS finds a much lower effective
$\xi$ slope $(\sim 1.01)$ that is attributed to a significant
component of Reggeon (not Pomeron) exchange.

From the results of the fit in Refs.~\cite{H1,All} mesons trajectories
become less important at large $\beta$. If this is really the 
case, the $\xi$ slope for the Pomeron should increase when
going to smaller $\beta$ values. 

A $\beta$ independent $\xi$-slope, when only the Pomeron is 
exchanged, would be a quite natural result but other possibilities
are not excluded. In order to clarify better this point 
we consider the data of Ref.~\cite{TA} for $F_2^{D(3)}$ and
parametrize the structure function, for each fixed $\beta$
value, with the simple function
\begin{equation}
F_2^{D(3)}(\beta_i,Q^2,\xi)\simeq (d_0+d_1 \ln(Q^2/Q_0^2))
\xi^{-n(\beta_i)}
\label{z7}
\end{equation}
where $\beta_i=0.065, 0.175, 0.375, 0.65$.

A FORTRAN program for function minimization and error analysis
(MINUIT) determine the parameters, in particular $n(\beta_i)$.
The output of the program has been plotted in Fig.2, the open
squares, and agrees with the result of Ref.~\cite{H1} within
the errors. When the cut $\xi\leq 0.005$ is applied to the data 
and the minimization procedure is repeated, $n(\beta)$ 
changes to the values indicated with full triangles in Fig.2.
Results are not reported at $\beta=0.065$, since the cut 
leaves only two points, and at $\beta=0.65$ since practically
all the points satisfy the selection criterion. Other 
parametrizations for $F_2^{D(3)}$, see for 
example Refs.~\cite{noi1,noi2}, reproduce the same trend for $n(\beta)$.

While this rather crude analysis cannot be taken too
seriously, more intriguing is the prediction of Ref.~\cite{GNZ}
for $n(\beta)$ shown as a dashed line in Fig.2 and evaluated
as follows. We first calculate
\begin{equation}
n(\xi,\beta)=\frac{\partial \ln F_2^{D(3)}(\xi,\beta)}{\partial
\ln(1/\xi)}
\label{z8}
\end{equation}
following the method of Ref.~\cite{GNZ}, with the values for the 
parameters appropriate to small $\xi$, and then take the mean
value of $n(\xi,\beta)$ in the range $0.0003\leq \xi \leq 0.005$:
\begin{equation}
n(\beta)=\frac{1}{\Delta \xi}\int_{\xi_L=0.0003}^{\xi_U=0.005}
n(\xi,\beta)\,d\xi~,
\label{z9}
\end{equation}
where $\Delta \xi=\xi_U-\xi_L$. When averaged over $\xi$, the
increase of $n(\beta)$, when $\beta$ decreases, becomes very
slow and finally $n(\beta)$ should tend to a constant in the
dipole Pomeron model ~\cite{GNZ}.

In order to understand better this effect we will consider, 
in the next Section, the problem starting from Eq.~(\ref{z6}) 
and a Regge model for the Pomeron flux in the 
proton ~\cite{noi1,noi2}.

{\bf 3. The t-dependence of $F_2^{D(4)}$ and factorization}

We consider a simplified model for the Pomeron flux, defined as
\begin{displaymath}
F_{P/p}(\xi,t)=\xi |\Phi(\xi,t)|^2~,
\end{displaymath}
where a linear Pomeron trajectory $\alpha(t)=\alpha(0)+\alpha 't$
gives ~\cite{noi1,noi2}
\begin{equation}
F_{P/p}(\xi,t)=c \xi^{1-2\alpha(0)}e^{2\alpha '(B-\ln(\xi))t}~.
\label{z10}
\end{equation}
The following considerations
apply also to a non-linear Pomeron trajectory ~\cite{noi2} with
obvious modifications.

Experimental data ~\cite{TA,MD} are given for
\begin{equation}
F_2^{D(3)}(\beta,Q^2,\xi)=\int_{t_+}^{t_-} F_{P/p}(\xi,t)
G_{q/P}(\beta,Q^2,t)~,
\label{z11}
\end{equation}
where $t_{\pm}$ are given for example in Ref.~\cite{noi2} and
$t_-\sim -m_p^2\xi^2 \gg t_+$, $m_p$ being the proton mass.

The assumption that the structure function $G_{q/P}$ does not
depend on the Pomeron invariant mass $t$ is made in 
Ref.~\cite{IS} in analogy with the structure function of the
photon and is considered there a "reasonable first approximation".
If $t$ is neglected in $G_{q/P}$ the rapidity gap variable 
$\xi$ is not correlated any more with $\beta$ and $Q^2$. 

In general, however, the vertex $\gamma^* -P-X$, the large 
black dot in Fig.1, will depend on $t$ and this dependence is,
per se, an interesting problem subject of numerous debates.
A source of uncertainty comes from the parametrization of the
triple Pomeron vertex (see e.g. Ref.~\cite{AK} and earlier
references therein). The most recent triple Pomeron fits were made as 
early as the mid'70-ies and neither of these fits accounts for
the rising cross sections, i.e. they involve a simple, unit
intercept Pomeron pole, inadequate from the present point of
view.

The effect we are studying is very small indeed and should be
visible at small $\beta$, that is in the triple Regge region.
A detailed analysis would require a refitting of the 
diffractive hadronic reactions with a modified Pomeron
trajectory and a larger sample of data with smaller errors.
For the time being we content ourselves to add another
argument, in a model calculation, in favour of the behaviour
for $n(\beta)$ predicted in Ref.~\cite{GNZ}. 

The physical mechanism at the basis of the two approaches is 
quite different but, 
as far as $n(\beta)$ is concerned, the result will be the same.
In practice, a meaningful comparison should consider only
the slope $dn(\beta)/d\beta$ of $n(\beta)$. If the two
results are to be consistent, $dn(\beta)/d\beta$ must
result negative and small in both calculations.
In appendix we give an estimate of the integral over $t$
of $F_2^{D(4)}$ that can be used if a series expansion of
$G_{q/P}$ near $t=0$ is known. Setting
\begin{equation}
r(\beta) \equiv \frac{a_1(\beta)}{2\alpha'a_0(\beta)}~,
\label{z12}
\end{equation}
where $a_0(\beta)$ and $a_1(\beta)$ are the first two coefficients of the 
series expansion (see Eq.~(\ref{A.1}), we get from Eq.~(\ref{A.3})
\begin{equation}
\frac{dn(\beta)}{d\beta}=-\frac{1}{\Delta\xi} \left(
\frac{dg(\xi_U)}{d\beta}-\frac{dg(\xi_L)}{d\beta} \right)~,
\label{z13}
\end{equation}
with $g(\xi)$ defined in Eq.~(\ref{A.4}). 
Taking into account that $\xi\leq 0.005$ and $B\simeq
7$ ~\cite{noi2}, $dg/d\beta$ can be easily obtained in the
convenient form ~\cite{BAT}
\begin{displaymath}
\frac{dg(\xi)}{d\beta}=r'(\beta) \frac{e^{\ln(\xi)-B}}{B-
\ln(\xi)+r(\beta)}\times
\end{displaymath}
\begin{equation}
\left(\sum_{m=1}^{M-1}\frac{(-1)^{m+1}\,m!}{(B-\ln(\xi)+
r(\beta))^m}+ O(|B-\ln\xi+r(\beta)|^M) \right)
\label{z14}
\end{equation}
that is a continuous and slowly decreasing function of $\xi$.

The sign of $r'(\beta)$ can be inferred as follows. We consider
again Ref.~\cite{DL} and, in this scheme, identify the
coefficients $a_0(\beta)$ and $a_1(\beta)$. 
The most important contribution to $a_o(\beta)$
will come from the box diagram, $G_{q/P}^a$ in Ref.~\cite{DL},
that does not depend on $t$, while $G_{q/P}^b$ (with an
appreciable t-dependence) will contribute mainly to $a_1(\beta)$.
An explicit form for $G_{q/P}^b$ is given only for $|t|>0.2\,GeV^2$,
hence, near $t=0$, care must be taken in parametrizing the data
for the triple Pomeron contribution to the inclusive 
differential cross section for the diffraction dissociation.
If the $\beta$ dependence is taken as in Ref.~\cite{DL}, it
turns out that $r(\beta)$ in Eq.~(\ref{z12}) is negative with 
a small positive derivative. Hence $n(\beta)$ according to Eq.~(\ref{z13})
is a decreasing function of $\beta$ and the prediction
of the Regge model approaches the result of the dipole
BFKL Pomeron. When other parameters, like $\alpha'\simeq 
0.25\,GeV^{-2}$, are chosen in a standard way, the two
results agree also from a quantitative point of view.

{\bf 4. Conclusions}

In this paper we studied the problem of gap factorization
for the Pomeron exchange. Once meson trajectories have been
suppressed, by considering values of $\xi$ such that
$\xi\leq 0.005$, only the Pomeron contributions remains.

We present first an analysis of experimental data ~\cite{TA,MD}
and show that the data sample, with a cut in the variable 
$\xi$, is compatible with a breaking of the gap factorization.
In the diffractive structure function $F_2^{D(3)}(\beta,Q^2,\xi)$
the $\xi$-dependence allows for an effective power $n(\beta)$ that 
depends on $\beta$, having a behaviour quite
different from the one predicted in Ref.~\cite{H1} where the
breaking comes from the interference of the Pomeron with the
exchanged meson trajectories.

We then evaluate $n(\beta)$ in a theoretical model ~\cite{GNZ}
and find a good agreement with our phenomenological analysis.
In this model the gap factorization is broken by the 
presence of two structure functions in the Pomeron, associated
with two different fluxes of Pomerons in the proton.

In a first instance it seems that in a Regge model 
~\cite{IS,DL,noi1,noi2,KG,GB,CAP}, the unique structure of the 
proton-Pomeron vertex implies the Pomeron factorization property.
The main result of this paper is the finding that the
t-dependence of the photon-Pomeron vertex may bias the 
commonly assumed Pomeron factorization. We show in fact
that the integration of $F_2^D(4)$ over $t$, in order to
obtain $F_2^{D(3)}$, can lead to an effective $\xi$-power 
$n(\beta)$, with the same behaviour as in Ref.~\cite{GNZ}.
For this purpose we use the model of Ref.~\cite{DL} and an 
approximate estimate of the $t$-integral, but we argue
that this effect is really present in any model.

As it appears in Fig.2, the breaking, proportional to the
slope of the dashed line, is very small and very difficult
to detect experimentally. New unpublished H1 data ~\cite{H1}
could already put limits on this breaking and further
elucidate the important question of the $t$-dependence in
the photon-Pomeron vertex. The knowledge of the variation
of this $t$-dependence with $\beta$ could be important in
clarifying the dynamics underlying the onset of the
triple Pomeron vertex.

\vskip 1.5cm
\underline{Acknowledgements}: The authors have profited from
conversations with E. Predazzi. One of us (L.L.J.) is grateful
to the Dipartimento di Fisica dell'Universita' di Padova,
to the Dipartimento di Fisica dell'Universita' della 
Calabria, and to the Istituto Nazionale di Fisica Nucleare, Sezione
di Padova and Gruppo Collegato di Cosenza for their
warm hospitality and financial support while part of this
work was done.

\vskip 0.5cm

{\bf Appendix}
\renewcommand{\theequation}{A.\arabic{equation}}
\setcounter{equation}{0}

Integrating Eq.~(\ref{z11}) over $t$ we keep the contribution of
the end point $t=t_-$ by considering $2\alpha'(B-\ln\xi)$ as a 
large parameter. According to the Laplace method ~\cite{EVG}
\begin{displaymath}
\int_{t_+}^{t_-} e^{2\alpha'(B-\ln\xi)t} G_{q/P}(\beta,Q^2,t)~
dt \sim
\end{displaymath}
\begin{equation}
e^{2\alpha'(B-\ln\xi)t_-}\sum_0^{\infty}\frac{a_n(\beta,
Q^2)}{[2\alpha'(B-\ln\xi)]^{n+1}},
\end{equation}
where
\begin{displaymath}
a_n(\beta,Q^2)=\left( \frac{\partial G_{q/P}(\beta,Q^2,
t_--u)}{\partial u} \right)^{(n)}_{u=0}.
\end{displaymath}
The estimate (\ref{A.1}) can be useful if the coefficients $a_n$
tend to zero rapidly enough or, better, if the series can be
truncated. The determination of the coefficients
$a_n$ will be possible when more accurate data will be available.
The saddle point at $t^*$, where $t^*$ is the solution of the 
equation
\begin{displaymath}
-2\alpha'(B-\ln\xi)+\frac{\partial \ln G_{q/P}}{\partial |t|} = 0~,
\end{displaymath}
does not contribute if $G_{q/P}$ is a decreasing function of $|t|$.

As an example, suppose to truncate the series in Eq.~(\ref{A.1}) to the
first two terms. Then Eq.~(\ref{z8}) gives
\begin{equation}
n(\xi,\beta)=2\alpha(0)-1+2\alpha't_--\frac{2}{B-\ln\xi}+
\frac{1}{B-\ln\xi+\frac{a_1(\beta)}{2\alpha'a_0(\beta)}}
\end{equation}
where no $Q^2$ dependence appears in accordance with the
experimental data ~\cite{H1}. The mean value of $n(\xi,\beta)$ can
be evaluated from Eq.~(\ref{z9}):
\begin{equation}
n(\beta)=2\alpha(0)-1-\frac{1}{\Delta\xi}[g(\xi_U)-g(\xi_L)]~,
\end{equation}
where the term
\begin{displaymath}
\frac{2\alpha'}{\Delta\xi}\int_{\xi_L}^{\xi_U}\,t_-\,d\xi
\sim \frac{2\alpha'}{3}(\xi_U^2+\xi_U\xi_L+\xi_L^2)
\end{displaymath}
has been neglected and
\begin{displaymath}
g(\xi) =
\end{displaymath}
\begin{equation}
e^B (2E_1[\ln(1/\xi)+B]-e^{a_1(\beta)/(2\alpha'a_0(\beta))}
E_1[\ln(1/\xi)+B+a_1(\beta)/(2\alpha'a_0(\beta)])~.
\end{equation}
Here $E_1(z)$ is the exponential integral \cite{BAT}


\newpage

\begin{thebibliography}{99}

\bibitem{CSS}
J.C. Collins, D.E. Soper and G. Sterman, Nuc. Phys. B {\bf 261},104 (1985);
G. Bodwin, Phys. Rev. D {\bf31}, 2616 (1985).

\bibitem{FS}
L. Frankfurt and M. Strikman, Phys. Rev. Lett. {\bf 64}, 1914 (1989).

\bibitem{CFS}
J.C. Collins, L. Frankfurt and M. Strikman, Phys. Lett. B {\bf 307}, 
161 (1993).

\bibitem{IS}
G. Ingelman and P. Schlein, Phys. Lett. B {\bf 152}, 256 (1985).

\bibitem{TA}
T. Ahmed et al., Phys. Lett. B{\bf 348}, 681 (1995).

\bibitem{MD}
M. Derrick et al., Z. Phys. {\bf C68}, 569 (1995).

\bibitem{KG}
K. Goulianos, hep-ex/9708004, August 1997.

\bibitem{H1}
H1 Collaboration, Contribution to ICHEP '96, Warsaw, Poland, 
July 1996, pa 02-061.
C. Adloff et al., H1 Collaboration, Z. Phys. {\bf C76}, 613 (1997).

\bibitem{H2}
H1 Collaboration, Contribution to HEP '97, Jerusalem, Israel,
August 1997, Abstract 377.


\bibitem{DL}
A. Donnachie and P.V. Landshoff, Phys. Lett. {\bf 191}, 309 (1987),
Nuc. Phys. B {\bf 303}, 634 (1988).

\bibitem{All}
A. Metha, in Proceedings Hard Diffractive Scattering, ed. by
H. Abramowicz et al. (Tel Aviv, 1996), 710;
K. Golec-Biernat and J. Kwiecinski, Phys. Rev. D {\bf 55}, 3209 (1997);
R. Fiore et al., Proceedings of Diquark 3, Torino, October 1996,
to appear.

\bibitem{GNZ}
M. Genovese, N.N. Nikolaev and B.G. Zakharov, JETP {\bf 81}, 625
(1995).

\bibitem{NZ}
M. Genovese, N.N. Nikolaev and B.G. Zakharov, Phys. Lett. B 
{\bf 380}, 213 (1996), Phys. Lett. B {\bf 378},347 (1996);
N.N. Nikolaev, W. Sch\"afer and B.G. Zakharov, preprint 
KFA-IKP (Th)-1996-06, July 1996.

\bibitem{BFKL}
E.A. Kuraev, ,L.N. Lipatov and V.S. Fadin, JETP {\bf 45}, 199
(1977); Ya.Ya. Balitskii and L.N. Lipatov, Sov. J. Nucl.
Phys. {\bf 28}, 822 (1978).

\bibitem{WU}
M. W\"usthoff, Phys. Rev. D {\bf 56}, 4311 (1997).

\bibitem{GB}
K. Golec-Biernat and J. Kwiecinski, Phys.Lett. {\bf B353},
329 (1995).

\bibitem{CAP}
A. Capella et al., Phys. Lett. {\bf B343}, 403 (1995).

\bibitem{ZEUS}
ZEUS Collaboration, preprint DESY 97-184.

\bibitem{noi1}
R. Fiore, L.L. Jenkovszky and F. Paccanoni, Phys. Rev. D {\bf 52},
6278 (1995).

\bibitem{noi2}
R. Fiore, L.L. Jenkovszky and F. Paccanoni, Phys. Rev. D {\bf 54},
6651 (1996).

\bibitem{AK}
A. Kaidalov, Phys. Rep. {\bf 50}, 157 (1979);
K. Goulianos, Phys. Rep. {\bf 101}, 169 (1983).

\bibitem{BAT}
Higher Transcendental Functions (Bateman Manuscript Project)
edited by A. Erdelyi et al. (Mc Graw-Hill, New York, 1953) Vol. II.

\bibitem{EVG}
M.A. Evgrafov, Asymptotic estimates end entire functions,
(Gordon and Breach, New York, 1961).


\end{thebibliography}
\end{document}